\documentstyle[epsfig,amsmath,amssymb,graphicx,soul,color,subfigure,lscape]{jaa}

\begin{document}



\title[Critical Assessment of Flux Transport Dynamo]{A Critical Assessment of the Flux Transport Dynamo}
\author[Arnab Rai Choudhuri]{Arnab Rai Choudhuri\\
Department of Physics\\
Indian Institute of Science\\
Bangalore - 560012. India}

\pubyear{2014}
\volume{xx}
\date{Received xxx; accepted xxx}

\maketitle

\label{firstpage}

\begin{abstract}

We first discuss how the flux transport dynamo with reasonably high diffusion can
explain both the regular and the irregular features of the solar cycle quite well.
Then we critically examine the inadequacies of the model and the challenge posed by
some recent observational data about meridional circulation, arriving at the conclusion
that this model can still work within the bounds of observational data.
 
\end{abstract}


\section{Introduction}

The flux transport dynamo model has emerged in recent years as the most promising 
theoretical model for explaining different aspects of the solar cycle. While 
the flux transport dynamo model
may not yet be unanimously accepted in the solar physics community and
doubts are often raised about its validity, no other alternative model of the
solar cycle has been studied to the same depth. Following an early paper by
Wang, Sheeley \& Nash (1991) proposing the basic idea, the first two-dimensional
models of the flux transport dynamo were constructed by Choudhuri, Sch\"ussler \&
Dikpati (1995) and Durney (1995).  From 1995 onwards, about a dozen papers by different
groups on the flux transport dynamo have by now received more than 100 citations
according to ADS (Choudhuri, Sch\"ussler \& Dikpati 1995; Durney 1995; Dikpati
\& Charbonneau 1999; Charbonneau \& Dikpati 2000; K\"uker, R\"udiger \& Shultz
2001; Dikpati \& Gilman 2001; Nandy \& Choudhuri 2002; Dikpati et al.\ 2004;
Chatterjee, Nandy \& Choudhuri 2004; Dikpati, de Toma \& Gilman 2006; Dikpati
\& Gilman 2006; Choudhuri, Chatterjee \& Jiang 2007). During the same period,
we are aware of only three papers dealing with alternate models of the solar dynamo
which received more than 100 citations (R\"udiger \& Brandenburg 1995; Charbonneau
\& MacGregor 1997; Beer, Tobias \& Weiss 1998).  Interestingly, two of the
authors of these papers dealing with alternate models (R\"udiger, Charbonneau)
later became votaries of the flux transport dynamo and wrote important papers
on this subject (Dikpati \& Charbonneau 1999; K\"uker, R\"udiger \& Shultz 2001).  
This simple consideration of the citations data makes it amply
clear that the flux transport dynamo model has received much more attention in
recent years than any alternate model of the solar cycle.  If the flux 
transport dynamo model proves to be incorrect, then we shall be left with no
other theoretical model that can explain various aspects of the solar cycle
in such detail.  It is therefore important to critically assess the flux
transport dynamo model and to examine if the doubts and uncertainties about
this model are serious enough.  This is attempted here.

\section{Basics of the flux transport dynamo}

The basic idea of solar dynamo theory is that the toroidal and the poloidal
components of the Sun's magnetic field sustain each other through a feedback loop.
It is easy to see that the poloidal magnetic field can be stretched by
differential rotation to generate the toroidal magnetic field.  Since the
differential rotation of the Sun has now been mapped by helioseismology, this
is now a well-understood process.  

The complementary process of generation of the poloidal field from the toroidal
field is less well established. There have been historically two schools of thought.
Parker (1955) suggested and then Steenbeck, Krause \& R\"adler (1966) elaborated
that the helical turbulence in the solar convection zone can twist the toroidal
field to give rise to the poloidal field.  This process is called the $\alpha$-effect
and is possible only if the strength of the toroidal field is such that the magnetic
energy density is less than the energy density of turbulence.  It is estimated
that the toroidal field cannot be stronger than $10^4$ G in order to be twisted
by turbulence. The second school of thought is due to Babcock (1961) and 
Leighton (1964). Sunspot pairs forming out of the toroidal magnetic field
have tilts produced by the Coriolis force (D'Silva \& Choudhuri 1993)---tilts
increasing with latitude in accordance with Joy's law.  According to the
Babcock--Leighton viewpoint, the decay of the tilted sunspot pairs gives rise
to the poloidal magnetic field.

Early solar dynamo models in 1970s and 1980s usually assumed that the poloidal
field is generated by the $\alpha$-effect.  This assumption had to be questioned
when magnetic buoyancy calculations based on the thin flux tube equation (Spruit
1981; Choudhuri 1990) indicated that the toroidal field at the base of the Sun's
convection zone is as strong as $10^5$ G (Choudhuri \& Gilman 1987; Choudhuri 1989;
D'Silva \& Choudhuri 1993; Fan, Fisher \& DeLuca 1993). Weaker magnetic fields would
be affected by the Coriolis force in a way that is not consistent with observational
data.  Although there are some mechanisms of suppressing the Coriolis force (Choudhuri
\& D'Silva 1990; D'Silva \& Choudhuri 1991), these are not expected to be very effective
inside the solar convection zone. Since helical turbulence in the convection zone
will not be able to twist such a strong toroidal field, the Babcock--Leighton (BL)
mechanism seems to be the best option for generating the poloidal field.

One requirement of any dynamo model is that a dynamo wave has to propagate
equatorward, in order to explain the appearance of sunspots at lower latitudes
with the progress of the solar cycle. According to what is called the Parker--Yoshimura
sign rule (Parker 1955; Yoshimura 1975), the $\alpha$-coefficient and the differential
rotation $\Omega (r, \theta)$ have to satisfy the following condition in the northern
hemisphere
$$\alpha \frac {\partial \Omega}{\partial r} < 0 \eqno(1)$$
in order to produce equatorward propagation. Now, the BL mechanism can also be
mathematically represented by a coefficient $\alpha$ appearing in the equations
exactly like the $\alpha$-effect.  It is clear from the observations of sunspot
tilts that the $\alpha$ corresponding to the BL mechanism has to be positive in
the northern hemisphere. Since helioseismology finds $\partial \Omega/ \partial r$
to be positive in the lower latitudes where sunspots are seen, it appears that
condition (1) is not satisfied and a dynamo model in which the poloidal field
is generated by the BL mechanism may not reproduce solar-like behaviour.

The Parker--Yoshimura sign rule (1) was derived without considering the effects
of the meridional circulation.  It is known that the meridional circulation
of the Sun advects the poloidal field poleward near the solar surface (Wang, Nash \&
Sheeley 1989; Dikpati \& Choudhuri 1995; Choudhuri \& Dikpati 1999). Although the
meridional circulation is poleward just below the Sun's surface, its nature deeper down
is still not unanimously established from observational data.  Since this circulation
is driven by the turbulent stresses in the convection zone, the meridional circulation is expected on theoretical
grounds to be confined within the convection zone. The simplest assumption is that
of an equatorward return flow at the bottom of the convection zone.  Choudhuri,
Sch\"ussler \& Dikpati (1995) showed that an equatorward meridional
circulation at the bottom of the convection zone can overrule the condition (1)
and can make the toroidal field produced there shift equatorward with time, reproducing
the solar behaviour.

\begin{figure}
\center
\includegraphics[width=6cm]{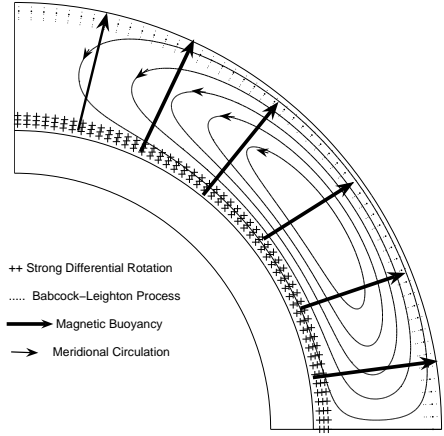}
  \caption{A cartoon explaining how the flux transport dynamo works.}
\end{figure}

Figure 1 is a cartoon representation of the flux transport dynamo. The tachocline
at the bottom of the convection zone is the region of concentrated differential rotation
where the strong toroidal field is produced.  This toroidal field rises to the solar
surface due to magnetic buoyancy to produce sunspots.  At the surface the BL mechanism 
acts on the decaying sunspots to generate the poloidal
field.  This poloidal field is carried poleward by the meridional circulation
as seen in the observational data.
Ultimately the poloidal field has to be brought to the bottom of the convection
zone, where the strong differential rotation can act on it. We shall make some
comments on the transport of the poloidal field in the next section. Apart from
modelling the solar cycle, such flux transport dynamo models are now being used
to model different aspects of stellar cycles (Jouve, Brown \& Brun 2010; Karak,
Kitchatinov \& Choudhuri 2014).

\section{Two types of dynamo and modelling irregularities}

\def\tc{\tau_{\rm conv}}
\def\tt{\tau_{\rm tach}}
\def\ta{\tau_{\rm adv}}

As mentioned in the previous section, the poloidal field generated by the BL mechanism
near the solar surface has to be transported to the bottom of the convection zone in
order for the dynamo to operate. The two obvious ways of doing this is through diffusion
and through advection by the meridional circulation.  Let us look at these possibilities.
The turbulent diffusion within the body of the convection zone is expected to be much
stronger than that within the tachocline.  Hence the diffusion time $\tc$ within the
convection zone has to be much shorter than the diffusion time $\tt$ within the tachocline.
Now let $\ta$ be the advection time by meridional circulation.  It is necessary
for $\ta$ to be shorter than $\tt$ in order for the equatorward propagation of the
toroidal field at the bottom of the convection zone, overcoming the condition (1).
Now the conditions $\tc < \tt$ and $\ta < \tt$ can be satisfied in two possible ways:
$$\tc < \ta <\tt \eqno(2)$$
or
$$\ta < \tc <\tt. \eqno(3)$$
There have been two types of solar dynamo models corresponding to these two possibilities.
The dynamo model developed by Choudhuri and his collaborations (Nandy, Chatterjee, Jiang, Karak) 
satisfy (2) and is known as the high-diffusion or diffusion-dominated model, in which the
transport of the poloidal field takes place primarily due to diffusion. On the other hand,
the dynamo model developed by Dikpati and her collaborations (Charbonneau, Gilman, de Toma) 
satisfy (3) and is known as the low-diffusion or advection-dominated model, in which the
transport of the poloidal field takes place primarily due to advection by meridional circulation. The differences
between these two types of models have been systematically studied by Jiang, Chatterjee \&
Choudhuri (2007) and Yeates, Nandy \& Mackay (2008).

Both types of solar dynamo model have been able to reproduce various regular aspects of
the solar cycle.  However, the high-diffusion model satisfying (2) succeeds better in
explaining the dipolar parity of the Sun (Chatterjee, Nandy \& Choudhuri 2004; Hotta \&
Yokoyama 2010) or the lack of significant hemispheric asymmetry (Chatterjee \& Choudhuri
2006; Goel \& Choudhuri 2009). After explaining the regular aspects of the solar cycle, the
thrust of research in the last few years has been to explain the irregularities of the solar
cycle. It appears that the high-diffusion model satisfying (2) gives much better agreement
with observations when modelling the irregularities of the solar cycle.  This is a vast
subject.  We make only a few general remarks.  The readers are referred to a recent review
(Choudhuri 2014) for a more detailed discussion.

It has been known for a while that fluctuations in the poloidal field generation mechanism can
produce irregularities in the solar cycle (Choudhuri 1992).  Within the framework of the
flux transport dynamo, Choudhuri, Chatterjee \& Jiang (2007) suggested how such fluctuations
arise. The BL mechanism for the poloidal field generation depends on the tilts of bipolar
sunspots.  One observationally finds a scatter in the tilts around the average given by
Joy's law---presumably produced by the buffeting of rising flux tubes by convective
turbulence (Longcope \& Choudhuri 2002).  This introduces a randomness in the BL mechanism,
for which we now have strong observational support (Dasi-Espuig et al.\ 2010; Kitchatinov
\& Olemskoy 2011). If the irregularities of the sunspot cycle arise in this way, then 
Jiang, Chatterjee \& Choudhuri (2007) found that the high-diffusion model can explain
the observed correlation between the polar field at the sunspot minimum and the strength
of the next cycle---a correlation which the low-diffusion model cannot reproduce.  It may
be noted that at the end of cycle 23 there were theoretical efforts in predicting the strength
of cycle 24 on the basis of dynamo models. On incorporating the data corresponding to the
weak polar field at the end of cycle 23, Choudhuri, Chatterjee \& Jiang (2007) found that
cycle 24 comes out as a weak cycle in the high-diffusion model---which is expected on the
basis of the correlation found in this model.  However, Dikpati \& Gilman (2006) predicted
from  their low-diffusion model that cycle 24 would be exceptionally strong. By now there
is enough evidence that the prediction of the high-diffusion model is much closer to the
truth, as seen in Figure~2.

\begin{figure}
\center
\includegraphics[width=8cm]{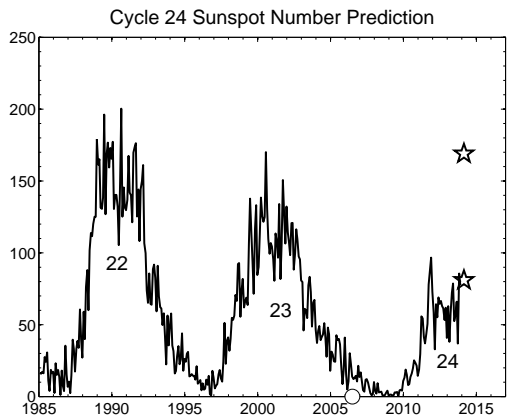}
  \caption{The monthly sunspot number plot for the last few years, indicating the
theoretical predictions. The upper star is the peak of cycle~24 predicted by
Dikpati and Gilman (2006), whereas the lower star is what was predicted by Choudhuri,
Chatterjee and Jiang (2007). The circle on the horizontal axis indicates the time when
these predictions were made (in 2006).}
\end{figure}

Since the strength of the meridional circulation sets the period of the dynamo (Dikpati \&
Charbonneau 1999), it is no wonder that variations in the meridional circulation also
produce irregularities in solar cycles.  Karak \& Choudhuri (2011) found that such fluctuations
introduced in the high-diffusion model can explain the Waldmeier effect, which the low-diffusion
model cannot explain at all.  This provides another support in favour of the high-diffusion
model. Recently the high-diffusion model has been used to model the grand minima.  A grand
minimum can be caused either by the weakness of the poloidal field during the sunspot minimum
(Choudhuri \& Karak 2009; Olemskoy, Choudhuri \& Kitchatinov 2013) 
or by the weakness of the meridional circulation (Karak 2010).
On considering both these effects simultaneously, Choudhuri \& Karak (2012; see also
Karak \& Choudhuri 2013) found that they
can explain the statistics of occurrence of grand minima.

It should be clear from the above discussion that the high-diffusion model is a better
description of what is happening inside the solar convection zone.  Apart from the two
poloidal field transport mechanisms mentioned at the beginning of this section, a third
possible mechanism has been recognized recently: downward turbulent pumping (Karak \& Nandy 2012;
Jiang et al.\ 2013).  The effect of this is similar to high diffusion. On including downward
turbulent pumping in the low-diffusion model, the model starts behaving somewhat like
the high-diffusion model.

\section{Inadequacies of the present models}

In spite of the success in modelling different regular and irregular aspects of the
solar cycle, the flux transport dynamo model at present has several inadequacies.  We
now point out some of them.

Since the differential rotation within the tachocline is strongest in high latitudes,
there is a tendency of strong toroidal fields being produced in high latitudes (Dikpati
\& Charbonneau 1999; K\"uker, R\"udiger \& Schultz 2001).  The intriguing question is
why sunspots appear only at the lower latitudes.  One suggestion by Nandy \& Choudhuri (2002; see also Guerrero \& Mu\~noz 2004)
is that the meridional circulation penetrates slightly below the bottom of the convection
zone, ensuring that the toroidal field produced at the high latitudes is pushed into the
stable layers to avoid sunspot eruptions at high latitudes.  While this hypothesis has
been questioned by some authors (Gilman \& Miesch 2004), the fact that torsional oscillations
begin at high latitudes before the beginning of the sunspot cycle lends a strong support
to this hypothesis (Charkraborty, Choudhuri \& Chatterjee 2009). Some authors (Hotta \&
Yokoyama 2010; Mu\~noz-Jaramillo et al.\ 2010) artifically restricted the sunspot eruptions
to low latitudes by confining the $\alpha$-coefficient to low latitudes without providing
any physical justification for this.  We have to admit that at present there is no
concensus amongst dynamo theorists why sunspots do not appear at high latitudes.

Magnetic buoyancy and the BL mechanism, which are essential ingredients of the flux transport
dynamo, are inherently three-dimensional.  They can be treated only rather crudely in
two-dimensional models. Choudhuri, Nandy \& Chatterjee (2005) found that two widely used methods
for specifying magnetic buoyancy give quite different results when all the other things are
kept the same.  The best way of treating the BL mechanism in two dimensions is also debated.
Nandy \& Choudhuri (2001) concluded that using the $\alpha$-coefficient and using the double
ring procedure proposed by Durney (1995) give similar results.  However, recently Mu\~noz-Jaramillo
et al.\ (2010) have claimed that the double ring procedure is the superior procedure.
Perhaps the next step is to construct kinematic models in which magnetic buoyancy and
the BL mechanism are treated in three dimensions.  Yeates \& Mu\~noz-Jaramillo (2013) have
initiated such calculations.  It remains to be seen whether this approach reproduces
the results of two-dimensional models.

Since there are no sunspots during a grand minimum, we expect that the BL mechanism will not
be operative at that time.  If the BL mechanism is the only mechanism for generating the
poloidal field, then we do not understand how the dynamo comes out of a grand minimum.
Most probably we need something like the traditional $\alpha$-effect to pull the dynamo
out of a grand minimum (Karak \& Choudhuri 2013; Hazra, Passos \& Nandy 2014). Does this
$\alpha$-effect operate all the time along with the BL mechanism or does it become effective
only during grand minima when magnetic fields are weaker? At present, we have very little
understanding of these issues.

Finally, the usual assumption in any mean field theory like the solar dynamo theory is that
fluctuations have to be small compared to the mean fields.  Since the magnetic field exists
in the form of flux tubes within the convection zone, this is certainly not true.  How the
presence of flux tubes affects the mean field theory has still not been studied adequately
(Choudhuri 2003).  We certainly need to take account of the flux tubes in order to explain
certain aspects of the sunspot cycle.  For example, one mechanism for producing the observed helicity 
of active regions is that the poloidal field gets wrapped around rising flux tubes (Choudhuri
2003; Choudhuri, Chatterjee \& Nandy 2004; Chatterjee, Choudhuri \& Petrovay 2006; Hotta \&
Yokoyama 2012). Presumably, the mean field theory somehow captures the essence of magnetic
field dynamics even though the fluctuations are larger than the mean.

\section{Recent challenges to the flux transport dynamo}

Although the inadequacies of the present dynamo models described in the previous section
makes it clear that we still do not understand many aspects of these models, these
inadequacies do not pose a threat to the validity of the flux transport model itself.
We now come to some recent developments which raise questions whether the flux transport
dynamo model is the correct model for the solar cycle.

As indicated in Figure~1, flux transport dynamo models usually assume a single cell of
meridional circulation spanning one hemisphere of the convection zone.  There is enough
observational evidence for a poleward flow in the upper layers of the convection zone.
While the equatorward return flow in the lower layers is still not established by observational
data, such a flow is needed to overcome the Parker--Yoshimura sign rule (1) so that we get
solar-like behaviour (Choudhuri, Sch\"ussler \& Dikpati 1995).  Of late, several groups
have claimed to find evidence of the equatorward return flow around the middle of the
convection zone rather than its bottom (Hathaway 2012; Zhao et al.\ 2013; Schad, Timmer
\& Roth 2013). It is possible that there are additional cells of meridional circulation
below this return flow, although the presently available observational data are not capable of settling
this issue. The important question facing us now is whether the flux transport dynamo
model can work if the equatorward return flow is at the middle of the convection zone
rather than at its bottom where the toroidal field is produced by differential rotation.

Let us first consider the situation that there is shallow cell of meridional circulation
with the return flow at the middle of the convection zone, with no flows underneath.
Guerrero \& de Gouveia Dal Pino (2008) showed that a solar-like behaviour (i.e.\
butterfly diagrams corresponding to equatorward propagation) can still be
obtained if there is equatorward turbulent pumping.  Since the existence of equatorward
pumping is less well established than the existence of downward turbulent pumping, we address the
question whether we can obtain solar-like behaviour without such pumping. Jouve \& Brun
(2007) considered radially stacked multiple cells in which the flow at the bottom of the
convection zone was poleward and solar-like behaviour was not reproduced. Recently Hazra,
Karak \& Choudhuri (2014) found that solar-like behaviour can be obtained as long as
there is an equatorward flow at the bottom of the convection zone even if there are multiple
cells of meridional circulation with an equatorward return flow in the middle of the convection zone. Although the
flux transport dynamo model was historically developed by considering initially a single cell
of meridional circulation, it seems that the model can still work with more complicated 
meridional circulation as suggested by recent observations.  Only if future observations
show that there is no equatorward flow at the bottom of the convection zone, the validity
of the flux transport dynamo model will have to be questioned.

\section*{Acknowledgements}

I thank DST for partial support through a J C Bose Fellowship.

\def\apj{{\em ApJ}}
\def\aap{{\em A\&A}}
\def\sol{{\em Sol. Phys.}}
\def\mn{{\em MNRAS}}

\end{document}